\documentclass[aps,physrev,preprint,groupedaddress,amsmath,amssymb]{revtex4-2}

\usepackage[utf8]{inputenc}
\usepackage[english]{babel}
\usepackage{hyperref}
\usepackage[labelfont=bf]{caption}
\usepackage[labelformat=simple]{subcaption}

\usepackage[outercaption]{sidecap}
\usepackage{graphicx}
\usepackage{dcolumn}
\usepackage{bm}
\usepackage{siunitx}
\usepackage{mathtools}
\usepackage{adjustbox}
\usepackage{xcolor}

\begin{document}

\title{Quantifying Stress States of Theoretically Modelled Polarimetric Measurements on Dielectric Media}

\author{Felix B. Müller}
\email[]{felix.mueller@ifnano.de}
\affiliation{Department Photonic Sensor Technologies, Institut für Nanophotonik Göttingen, Hans-Adolf-Krebs-Weg 1, 37077 Göttingen, Germany}
\author{Georgios Ctistis}
\email[]{georgios.ctitis@ifnano.de}
\affiliation{Department Photonic Sensor Technologies, Institut für Nanophotonik Göttingen, Hans-Adolf-Krebs-Weg 1, 37077 Göttingen, Germany}

\date{\today}

\begin{abstract} 
This work introduces and characterizes a theoretical model of a reflective polarimetric measurement technique determining the surface stress of a dielectric material, e.g. glass.
We have developed a procedure to reconstruct the actual stress state
%, which is the orientation and value of the principal axes of stress, 
from calculated Stokes vector components that would appear as polarization signals in a measurement.
For the calculation, we consider a special geometry of the principle stress axis, where we chose the third principal axis corresponding to the z axis to be perpendicular
to the surface and relaxed to a zero value.
%Our new approach of reconstructing surface stress states from reflected polarization states embraces the determination of the reflection Müller matrix.
%With that and an initial Stokes vector, the resulting reflected Stokes vector is calculated to create a database.
A created database represents the dependence of the reflected Stokes vector on the initial Stokes vector and the stress states included in the reflection matrix, which is iteratively calculated for several stress components and its orientations.
Introducing a model for the dependency of the reflected Stokes vector components on the stress states and fitting it to the database results in a system of equations of those dependencies which are solved for the stress state components and orientation.
Finally, we found a theoretical determination accuracy of the model for surface stress magnitude to be of the order of a few \SI{}{\mega\pascal} and the orientation of the principle axes of stress can be determined with an accuracy of \ang{20}.
\end{abstract}

\maketitle

\section{Introduction
\label{sec:introduction}}
Glass has developed throughout its more than 5000-year manufacturing history to be a versatile high-tech material, which is used in a wide range of applications, such as smart device displays or safety glass in general \cite{Ballauff2013, Berenjian2017, Richet2021}.
Especially in the automotive sector an increasing amount of sensors is nowadays included in windshields, drastically increasing the complexity in production \cite{Jacoby2022}, as well as their costs.
A sufficient quality control to minimize the failure of expensive glass panes and to save resources and energy is of great interest.
In the production of safety glass, residual stresses are purposely introduced during the tempering process to increase their stability and control their behavior at rupture \cite{Chen2013}.
An important part of quality control in the automotive industry is the prevention of glass pane break due to unintended residual stress distributions.
Therefore, a quantitative, large-scale inline monitoring of the stress distribution is desirable.\\
The measurement of the stress distribution in glass panes is an indirect measurement and is based on the change of the light polarization when polarized light interacts with an anisotropic medium, where the anisotropy arises from stress. 
This is the well-known effect of photoelasticity, first discovered by Brewster in 1816 \cite{Brewster1816}.
Currently, there are two commercially available devices that exploit this effect to estimate the surface stress of glass.
On the one hand, the grazing angle surface polarimeter (GASP) measures the surface stress on the tin-doped side of uncoated thermally treated float glass \cite{GASP0920}. 
The tin-doped side emerges from the nature of the float glass production process, which is based on liquid glass floating on liquid tin.
That is why this well-tried method has the restriction to its application on the tin-doped side of that special type of glass.
Additionally, the exit and entrance prism of the GASP have to be in contact with the glass surface using an immersion fluid, which complicates the handling and the scan over the glass pane, especially if it is curved, which is typically the case for automotive glass.
On the other hand, the scattered light polariscope (SCALP) measures the depth-dependent polarization changes of the backscattered light of a laser beam with an oblique incidence and extrapolates the fitted behavior to the surface \cite{Aben2010}.
Although it is also a well-established and typically used method, the following problems arise for the operation on the production line.
First, the SCALP also includes an exit and entrance prism in contact by means of an immersion fluid, which causes the same problems as for the GASP.
Second, the resolution of the stress measured with this technique and the applied fitting procedure is not completely sufficient for all types of glasses, especially for those with low residual stresses, e.g. laminated glasses, as typically used for windshields.
Third, the spatial orientation of the stress state inside the glass pane is not evaluated by this method.\\
In this work, we have developed a theoretical model for an experimental technique that uses the polarization state of the reflected beam to determine the surface stress conditions of an arbitrary dielectric material, on the example of glass.\\
The paper is organized as follows. First, we will give an introduction to the model and geometric considerations before explaining the creation of the database. We then quantify the model and consider the accuracy of the model before showing a first experimental approach and concluding the paper with a short outlook.

\section{Introduction to the model-building procedure
\label{sec:intro_model_proc}}
The comparison of the polarization state of the reflected to the incoming light beam contains information about the stress-optical properties of the sample.
In order to reconstruct the stress state from a measured polarization state, we developed a formalism that can describe the dependence of the polarization state on the stress conditions.
The polarization state is represented by a Stokes vector in this theoretical work as well as in the experimental application.
A phenomenological fitting procedure applied to that dependency enables us to create a system of equations that are solvable for the stress state, starting from the measured polarization states.\\
The procedure of determining the stress state can be tested by evaluating the fitting parameters from a theoretical database, which includes the originating stress states and the expected Stokes vector components.
The database is generated by systematically calculating the expected polarization state of a measurement for several combinations of the stress state in the surface of the dielectric sample.\\
For each data set, the Stokes vector components of a certain polarization state can be calculated using the Müller-formalism \cite{MuellerMatrixInProceedings1948} for the modeled setup of polarization optics and an optically anisotropic reflecting sample.
There, the polarization states are represented by a Stokes vector, and the optical components, including the reflecting dielectric sample, are represented by a Müller matrix.
The reflection matrix can be calculated first in the Jones-formalism \cite{Jones1941} for the special geometry of the stressed dielectric sample based on the $2 \times 2$-matrix formalism for the general interface transition of light between two biaxial dielectric media, as shown by Abdulhalim~\cite{Abdulhalim1999}.
Afterwards, the reflection matrix can be transferred to the Müller-formalism using the transformation in \cite{Podraza2017-482}.
The work of Abdulhalim \cite{Abdulhalim1999} uses the $2 \times 2$-matrix formalism considering not only the transmission, as was the case for multilayer systems in former publications \cite{Yeh1882, Lien1990, Gu1993}, but also the reflection of light at an
interface.
The $2 \times 2$-matrix formalism is used instead of the original $4 \times 4$-matrix formalism, e.g. published by \cite{Teitler1970, Berreman1972, Yeh1979} to simplify the eigenvalue problems of those matrices, which especially improves the application on more complicated anisotropic geometries and for non-perpendicular incident angles.\\
In the following, we will apply the formalism of Abdulhalim \cite{Abdulhalim1999} to our geometry of stress-induced anisotropy. 
We introduce thereby a Müller-matrix representation of the reflecting sample.
The resulting data will be fitted and the dependency function will be evaluated on its applicability to quantify optical polarization measurements with the originating stress state of the dielectric sample.

\section{Müller-matrix of the reflecting model sample
\label{sec:m-matrix}}
For the calculation of the polarization states after the reflecting sample the Müller-Stokes-formalism is used considering the Stokes-vector of the incoming beam shaped by the Müller-matrices of the optical components as well as the Müller-matrix of the reflecting sample.
The sample's reflection Müller-matrix corresponds directly to the Fresnel-matrix in the Jones formalism.
Our approach to find the matrix representation for our reflection geometry is thereby inspired by Abdulhalim \cite{Abdulhalim1999}.

\subsection{Special geometry of a medium stressed along two directions in the $xy$-plane and relaxed in $z$-direction
\label{sec:special_stress_geometry}}
The reflecting medium is a non-magnetic, dielectric, and non-absorbing material, which shows an optical anisotropy due to certain stresses.
For the application of surface stress measurements, we simplified a general geometry considering a relaxed stress direction perpendicular to the surface.
This defines the third principle axis of stress $\sigma_3=0$ perpendicular to the $xy$-surface-plane.
\begin{figure}[h]
    \centering
    \includegraphics[width=0.3\textwidth]{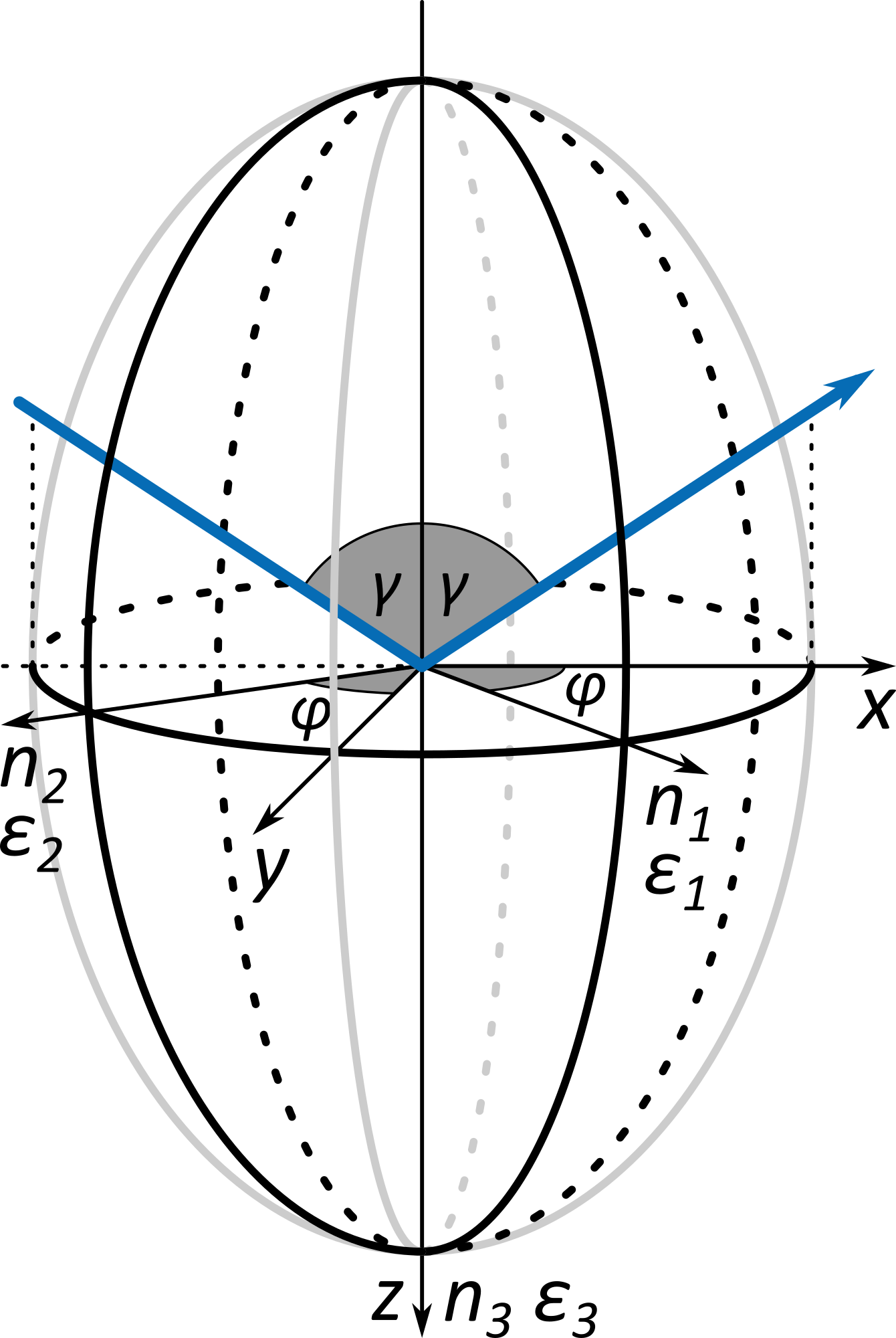}
    \caption{Index ellipsoid of an optically anisotropic medium with a special geometry of the anisotropy axes corresponding to the surface.
    The third principle axis of the refractive index $n_3$ or the permittivity $\varepsilon_3$, respectively, aligns with the $z$-axis perpendicular to the $xy$-surface plane, while $n_1$ and $n_2$ are arbitrarily rotated by $\varphi$ in the $xy$-surface-plane.
    The blue lines indicate the incoming and reflected beams, respectively, defining the plane of incidence in the $xz$-plane with the incidence angle $\gamma$.
    \label{fig:index_ellipsoid}}
\end{figure}
The resulting geometry is limited to scenarios of thermally tempered or float glass sufficiently far away from the edges, for which the absence of any stress component perpendicular to the surface is commonly assumed \cite{Thiele2022}.
Therefore, the proposed geometry is in any case suitable to apply the method to many commonly used glasses sufficiently far way from the edges.
However, the other generalities remain, resulting in the other two principle axes $\sigma_1$ and $\sigma_2$ being arbitrarily rotated by an azimuth angle $\varphi$ in the $xy$-plane.
Along those principal axes, the refractive indices $n_1$, $n_2$ and $n_3$ are present.
The geometry of the corresponding index ellipsoid is sketched in Fig.~\ref{fig:index_ellipsoid}.
For the depicted geometry the stress optical law simplifies to \cite{Aben1993}
\begin{subequations}
    \begin{eqnarray}
        n_1 &=& n_0 + C_1 \sigma_1 + C_2 \sigma_2 \\
        n_2 &=& n_0 + C_1 \sigma_2 + C_2 \sigma_1 \\
        n_3 &=& n_0 + C_2 \left( \sigma_1 + \sigma_2 \right),
    \end{eqnarray}
    \label{eq:StressOpticLawSimplified}
\end{subequations}
where $n_0$ is the refractive index of the material in a relaxed isotropic state, while $C_1$ and $C_2$ are the photoelastic constants.
This defines the diagonalized refractive index tensor corresponding to the principal axes coordinates.
For our calculations, we use $n_0=1.52$, measured for the soda-lime glass used in the intended experiments, as well as the parameter set $C_1=\SI{-0.65e-12}{\per\pascal}$ and $C_2=\SI{-4.22e-12}{\per\pascal}$ for vitreous silica, taken from Primak and Post \cite{Primak1959}.
In Cartesian coordinates the refractive index tensor has to be rotated from the main axes coordinates around the azimuth angle $\varphi$ using the rotation matrix $\mathbf{R}_z(\varphi)$ relative to the plane of incidence.
Moreover, the plane of incidence can also be rotated from the $xz$-plane around the $z$-axis by $\vartheta$.
That rotation is in general subtracted from the azimuth angle $\varphi$ regarding the $xyz$-coordinates resulting in the rotation matrix $\mathbf{R}_z(\varphi-\vartheta)$.
Figure~\ref{fig:index_ellipsoid} shows the case with $\vartheta=0$.
Thus, in the coordinates of the plane of incidence the permittivity tensor of a non-magnetic, dielectric, and non-absorbing material becomes \cite{Uranus2003}
\begin{widetext}
    \begin{eqnarray}
        \tensor{\varepsilon} &=& \tensor{n}^2\nonumber\\
        &=& \mathbf{R}(\varphi-\vartheta)\tensor{n}_\mathrm{diag} \mathbf{R}(-\varphi+\vartheta) \mathbf{R}(\varphi-\vartheta)\tensor{n}_\mathrm{diag} \mathbf{R}(-\varphi+\vartheta)\nonumber\\
        &=&
        \begin{pmatrix}
            \cos^2(\varphi-\vartheta) n_1^2 + \sin^2(\varphi-\vartheta) n_2^2 &\sin(\varphi-\vartheta)\cos(\varphi-\vartheta)(n_1^2 - n_2^2) &0\\
            \sin(\varphi-\vartheta)\cos(\varphi-\vartheta)(n_1^2 - n_2^2) &\cos^2(\varphi-\vartheta) n_2^2 + \sin^2(\varphi-\vartheta) n_1^2 &0\\
            0 &0 &n_3^2\\
        \end{pmatrix},        
        \label{eq:permittivityTensor}
    \end{eqnarray}
\end{widetext}
transforming the diagonal matrix in the principle axes coordinates by the rotation matrix around the coinciding $z$-axis.

\subsection{Reflection matrix of a biaxial medium with two principal axes arbitrarily rotated in the surface plane
\label{sec:reflMatrixSpecialGeometry}}
\begin{figure}[h]
    \centering
    \includegraphics[width=0.7\linewidth]{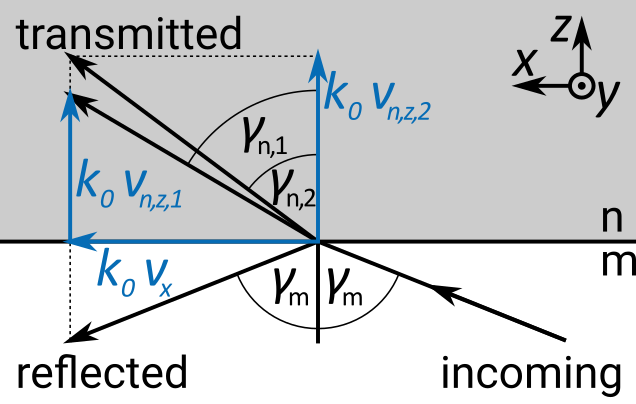}
    \caption{For the transition from an isotropic medium $m$ such as air or vacuum, into an anisotropic $n$ such as stressed glass, the reflection of a light beam consists of one mode while for the transmitted beam an additional mode appears due to birefringence.
    For both modes of the transmitted beam, the x-component of the $k$-vector stays the same as for the incoming beam, but the z-component differs for both modes.
    The sketch of the transition is a special case for $m$ isotropic and the plane of incidence coinciding with the $xz$-plane.
    The continuity and Maxwell's equations define the reflection and transmission coefficients in the Fresnel-matrices. The reflection matrix is transformed to the Müller-matrix-formalism in order to calculate the expected reflected Stokes vectors for a certain stress state.
    \label{fig:reflection_k}}
\end{figure}

The calculation of the reflection matrix for the special case of a biaxial medium with two principal axes arbitrarily rotated in the surface plane is calculated in the same way as shown for other cases by Abdulhalim in \cite{Abdulhalim1999}.
There, a certain condition for the electromagnetic fields of a light wave hitting an interface of, in general, two different optically anisotropic media is described.
From Abdulhalim \cite{Abdulhalim1999} the relation between the electric field state $\psi_e = (\sqrt{\varepsilon_0} E_x, \sqrt{\varepsilon_0} E_y)^T \exp{i k_z z}$ and the $2\times2$-matrix $\mathbf{G}$ is given by:
\begin{widetext}
    \begin{subequations}
        \begin{eqnarray}
            &\mathbf{G} \psi_e = 0
        	\label{eq:2x2_lin_diff_eq_short}\\
            &\mathbf{G} = 
            \begin{pmatrix}
                -(\varepsilon_{yx} + b_x \varepsilon_{yz} + b_x \nu_y \nu_z + \nu_x \nu_y)
                & \nu_x^2 + \nu_z^2 - b_y \nu_y \nu_z - b_y \varepsilon_{yz} - \varepsilon_{yy}\\
                \varepsilon_{xx} + b_x \varepsilon_{xz} + b_x \nu_x \nu_z - \nu_y^2 - \nu_z^2
                & \varepsilon_{xy} + b_y \varepsilon_{xz} + b_y \nu_x \nu_y + \nu_x \nu_y
            \end{pmatrix},
        	\label{eq:2x2_lin_diff_eq_gmatrix}
        \end{eqnarray}
        \label{eq:2x2_lin_diff_eq}
    \end{subequations}
\end{widetext}
where $\nu_p$ for $p \in \{x,y,z\}$ are the $k$-vector components normalised by $k_0$, $\varepsilon_{pq}$ for $p,q \in \{x,y,z\}$ are the components of the permittivity tensor $\tensor{\varepsilon}$ in Cartesian coordinates and
%\begin{equation}
%    b_{x/y} = \frac{\varepsilon_{zx/zy} + \nu_{x/y} \nu_z}{\nu_x^2 + \nu_y^2 - \varepsilon_{zz}}.
%    \label{eq:b_rel_psi_e_h}
%\end{equation}
\begin{subequations}
	\begin{eqnarray}
		b_{x} &=& \frac{\varepsilon_{zx} + \nu_{x} \nu_z}{\nu_x^2 + \nu_y^2 - \varepsilon_{zz}} \quad \textrm{and}	
    	\label{eq:b_rel_psi_e_h_x}\\
    	b_{y} &=& \frac{\varepsilon_{zy} + \nu_{y} \nu_z}{\nu_x^2 + \nu_y^2 - \varepsilon_{zz}}.
   		\label{eq:b_rel_psi_e_h_y}
	\end{eqnarray}
    \label{eq:b_rel_psi_e_h}
\end{subequations}
The magnetic field state $\psi_h = (\sqrt{\mu_0} H_x, \sqrt{\mu_0} H_y)^T \exp{i k_z z}$ corresponds to
\begin{subequations}
    \begin{eqnarray}
        &\psi_h = \mathbf{Q} \psi_e \\
        &\textrm{with} \quad
        \mathbf{Q} =
        \begin{pmatrix}
            b_x \nu_y &b_y \nu_y - \nu_z\\
            \nu_z - b_x \nu_x &-b_y \nu_x
        \end{pmatrix},
        \label{eq:rel_psi_e_h}
    \end{eqnarray}
\end{subequations}
in relation to the electric field \cite{Abdulhalim1999}.\\
The solutions $\nu_{zji}$ with $j \in \{1, 2, 3, 4\}$ and $i \in \{m, n\}$ of Eq.~\eqref{eq:2x2_lin_diff_eq_gmatrix} and the corresponding eigenvectors $V_{e/h,1/3,m/n}$ and $U_{e/h,2/4,m/n}$
for the particular medium and mode span the field vectors \cite{Abdulhalim1999}
\begin{subequations}
    \begin{eqnarray}
        \psi_{e,m/n}(r, t) &=&
        \begin{pmatrix}
            V_{e1,m/n} &U_{e2,m/n}
        \end{pmatrix}
        \begin{pmatrix}
            A_{1x}\\A_{2y}
        \end{pmatrix} \nonumber \\ && +
        \begin{pmatrix}
            V_{e3,m/n} &U_{e4,m/n}
        \end{pmatrix}
        \begin{pmatrix}
            A_{3x}\\A_{4y}
        \end{pmatrix} \\
        \label{eq:psi_e_with_modes} \textrm{and} \nonumber\\
        \psi_{h,m/n}(r, t) &=&
        \begin{pmatrix}
            V_{h1,m/n} &U_{e2,m/n}
        \end{pmatrix}
        \begin{pmatrix}
            A_{1x}\\A_{2y}
        \end{pmatrix} \nonumber \\ && +
        \begin{pmatrix}
            V_{h3,m/n} &U_{h4,m/n}
        \end{pmatrix}
        \begin{pmatrix}
            A_{3x}\\A_{4y}
        \end{pmatrix},
        \label{eq:psi_h_with_modes}
    \end{eqnarray}
    \label{eq:psi_with_modes}
\end{subequations}
where 
\begin{equation}
    V_{h,1/3} = \mathbf{Q} V_{e,1/3},\quad U_{h,2/4} = \mathbf{Q} U_{e,2/4}
    \label{eq:U-V-dependence}
\end{equation}
and
\begin{equation}
    A_{jl} = E_{jl} \exp (i k_0 (\nu_{jx} x + \nu_{jy} y + \nu_{jz} z) - i \omega t)
    \label{eq:eigenmodes_ampl}
\end{equation}
are the amplitudes of the eigenmodes with $j\in\{1,2,3,4\}$ and $l\in\{x,y\}$.
Using the  continuity conditions for the electroagnetic field, 
the Fresnel reflectivity matrix results in
\begin{widetext}
    \begin{equation}
        \mathbf{R}_{mn} = \left( \mathbf{E}_{12m}^{-1} \mathbf{E}_{34m} - \mathbf{H}_{12m}^{-1} \mathbf{H}_{34m} \right) \left( \mathbf{H}_{12n}^{-1} \mathbf{H}_{12m} - \mathbf{E}_{12n}^{-1} \mathbf{E}_{12m} \right),
        \label{eq:gen_refl-matrix_2x2}
    \end{equation}
\end{widetext}
with
\begin{subequations}
    \begin{eqnarray}
        \mathbf{E}_{ij,m/n} &=&
        \begin{pmatrix}
            V_{ei,m/n} &U_{ej,m/n}
        \end{pmatrix} \; \textrm{and}\\
        \mathbf{H}_{ij,m/n} &=&
        \begin{pmatrix}
            V_{hi,m/n} &U_{hj,m/n}
        \end{pmatrix}.
    \end{eqnarray}
    \label{eq:gen_refl_sub_matrix}
\end{subequations}
Considering the transition of an isotropic medium $m$, e.g. air or vacuum, to an anisotropic and biaxial dielectric material $n$, different modes have to be considered for the transmitted light wave in the anisotropic medium $n$ due to birefringence \cite{Abdulhalim1999}.
Figure~\ref{fig:reflection_k} schematically shows the situation for refraction.
In the isotropic medium $m$, the incoming and reflected waves are in the same mode for both directions of polarization.
The incoming wave mode can be described by its $k$-vector $\vec{k}_1m=\vec{k}_2m$ and the reflected by $\vec{k}_3m=\vec{k}_4m$.
Both waves are directed along the same angle $\gamma_m$ to the interface normal due to the law of reflection.
The transmitted wave is divided into two modes $\vec{k}_{1n}$ and $\vec{k}_{2n}$ with the corresponding angles $\gamma_{1,n}$ and $\gamma_{2,n}$ to the interface normal.
In that special case of materials and with the geometry described in Section~\ref{sec:special_stress_geometry} on page~\pageref{sec:special_stress_geometry}, a few simplifications directly arise.
Since the plane of incidence is in the $xz$-plane, the $y$-component of the wave vector $\vec{k}/k_0$ becomes $\nu_y = 0$.
Even if the plane of incidence is rotated by $\vartheta$ around the z axes in Fig.~\ref{fig:index_ellipsoid}, it can be considered as a additional rotation to the azimuth $\varphi-\vartheta$ instead of an additional $y$-component.
Equation~\eqref{eq:permittivityTensor}, describing the geometry of the permittivity,  defines $\varepsilon_{xz} = \varepsilon_{zx} = \varepsilon_{yz} = \varepsilon_{zy} = 0$ and $\varepsilon_{xy} = \varepsilon_{yx}$.
With those simplifications, in Eq.~\eqref{eq:b_rel_psi_e_h_y}, it directly follows that $b_y = 0$.
Hence, Eqs.~\eqref{eq:2x2_lin_diff_eq_gmatrix} and~\eqref{eq:b_rel_psi_e_h_x} result in
\begin{subequations}
    \begin{eqnarray}
        &\mathbf{G}_n =
        \begin{pmatrix}
            -\varepsilon_{xy} & \nu_{xn}^2 + \nu_{zn}^2 -\varepsilon_{yy}\\
            \varepsilon_{xx} + b_x \nu_{xn} \nu_{zn} - \nu_{zn}^2 & \varepsilon_{xy}
        \end{pmatrix} \quad\\
        &\textrm{with} \quad
        b_{xn} = \frac{\nu_{xn} \nu_{zn}}{\nu_{xn}^2 - \varepsilon_{zz}}
    \end{eqnarray}
    \label{eq:2x2_matrix_biaxial_simplified}
\end{subequations}
and Eq.~\eqref{eq:rel_psi_e_h} becomes
\begin{subequations}
    \begin{eqnarray}
        &\psi_h = \mathbf{Q} \psi_e \\
        &\textrm{with} \quad
        \mathbf{Q} =
        \begin{pmatrix}
            0 &-\nu_{zn}\\
            \nu_{zn} - b_{xn} \nu_{xn} &0
        \end{pmatrix}.
    \end{eqnarray}
    \label{eq:rel_psi_e_h_simplified}
\end{subequations}
The eigenvalues $\nu_{zn}$ as solutions of $\det(\mathbf{G}_n) = 0$ are presented in \cite{Abdulhalim1999} and simplify to
\begin{subequations}
    \begin{eqnarray}
        \nu_{z1n} &=& \sqrt{\frac{-b_n - \sqrt{b_n^2 - 4d_n}}{2}} \\
        \nu_{z2n} &=& \sqrt{\frac{-b_n + \sqrt{b_n^2 - 4d_n}}{2}} \\
        \textrm{with} \nonumber \\
        b_n &=& -\varepsilon_{xx} - \varepsilon_{yy} + \nu_{x,n}^2 \left(1 + \frac{\varepsilon_{xx}}{\varepsilon_{zz}} \right) \\
            d_n &=& \varepsilon_{xx} \varepsilon_{yy} - \varepsilon_{xy} \varepsilon_{yx} + \nu_{x,n}^2 \left( \frac{\varepsilon_{xy} \varepsilon_{yx}}{\varepsilon_{zz}} - \varepsilon_{xx} \right) \nonumber \\
            &&+ \frac{\nu_{x,n}^2}{\varepsilon_{zz}} (\nu_{x,n}^2 \varepsilon_{xx} - \varepsilon_{xx} \varepsilon_{yy})
        \\
        \nu_{x,n} &=& \nu_{x,m} = \sin{\gamma_m}.
    \end{eqnarray}
    \label{eq:k-components}
\end{subequations}
The eigenvectors should be determined under the consideration, that for the special case where the first principle axis is parallel to the $x$- and the second parallel to the $y$-direction, i.e. $\varepsilon_{xy} = \varepsilon_{yx} = 0$, the components $g_{11}$ and $g_{22}$ are not the denominator.
Furthermore, the convention ensures that the odd modes become p-polarised and the even modes s-polarised in that special case.
Considering those conditions, the relevant solutions of Eq.~\eqref{eq:2x2_matrix_biaxial_simplified} are
\begin{subequations}
    \begin{eqnarray}
        V_{e1,n} &=& N^{-1}
        \begin{pmatrix}
            1\\-\frac{g_{11}}{g_{12}}
        \end{pmatrix}
        \; \textrm{with} \; N = \sqrt{1 + \left( \frac{g_{11}}{g_{12}} \right)^2}
        \label{eq:2x2_eigenvector_V_biaxial}\\
        U_{e2,n} &=& M^{-1}
        \begin{pmatrix}
            -\frac{g_{22}}{g_{21}}\\1
        \end{pmatrix}
        \; \textrm{with} \; M = \sqrt{1 + \left( \frac{g_{22}}{g_{21}} \right)^2}\quad\quad
        \label{eq:2x2_eigenvector_U_biaxial}
    \end{eqnarray}
    \label{eq:2x2_eigenvector_V_U_biaxial}
\end{subequations}
in the biaxial medium $n$ with the third principle axis in $z$-direction.
For the isotropic medium $m$, e.g. air or vacuum, the $\mathbf{G}$-matrix in Eq.~\eqref{eq:2x2_lin_diff_eq} becomes
\begin{equation}
    \mathbf{G}_m =
    \begin{pmatrix}
        0 & \nu_{xm}^2 \nu_{zm}^2 -n_m^2\\
        n_m^2 + \nu_{zm}^2 \frac{\nu_{xm}^2 - 1}{\nu_{xm}^2 - n_m^2} & 0
    \end{pmatrix}.
    \label{eq:2x2_matrix_air_simplified}
\end{equation}
and the eigenvectors become
\begin{subequations}
    \begin{eqnarray}
        V_{e,1/3,m} &=&
        \begin{pmatrix}
            1 &0
        \end{pmatrix}^T
        \label{eq:2x2_eigenvector_V_air} \; \textrm{and}\\
        U_{e,2/4,m} &=&
        \begin{pmatrix}
            0 &1
        \end{pmatrix}^T
        \label{eq:2x2_eigenvector_U_air}
    \end{eqnarray}
    \label{eq:2x2_eigenvector_V_U_air}
\end{subequations}
for the eigenvalues $\nu_{z,1/2,m} = n_m \cos{\gamma_m}$ and $\nu_{z,3/4,m} = -n_m  \cos{\gamma_m}$ with $n_m = 1$ , which result from the $2\times2$-matrix formalism the same as expected from the sketch in Fig.~\ref{fig:reflection_k}.
Including these solutions in the Eqs.~\eqref{eq:psi_with_modes} to~\eqref{eq:gen_refl_sub_matrix} one gets the reflection Fresnel matrix $R_{mn}$ in the basis of the modes.
It can be transferred to the sp-basis by the transformation shown in \cite{Abdulhalim1999} equation~(22) and then translated to the reflection Jones matrix in the Müller-formalism \cite{MuellerMatrixInProceedings1948} as conventionally done, for example in \cite{Podraza2017-482}.\\
The resulting Müller-matrix $\mathbf{R}_\textrm{Müller}(\sigma_1, \sigma_2, \varphi)$ is used to perform step by step numerical calculations in order to predict measured data from the reflection matrix.
Those calculations start from Eq.~\eqref{eq:StressOpticLawSimplified} in order to determine the stress dependence of the reflection matrix $\mathbf{R}_{mn}(\sigma_1, \sigma_2, \varphi)$ in Eq.~\eqref{eq:gen_refl-matrix_2x2}.
The reflection Müller-matrix $\mathbf{R}_\textrm{Müller}(\sigma_1, \sigma_2, \varphi)$, given in the appended material in Eq.~\eqref{eq:jones_to_mueller}, directly results from the stress-dependent $\mathbf{R}_{mn}(\sigma_1, \sigma_2, \varphi)$ when one considers the solutions of the eigenvectors $V_{e/h,1/3,m/n}$ and $U_{e/h,2/4,m/n}$.
Their solutions for air in Eq.~\eqref{eq:2x2_eigenvector_V_U_air} lead to the electric and magnetic field matrices $\mathbf{E}_{ij,m}$ and $\mathbf{H}_{ij,m}$ for air in Eq.~\eqref{eq:refl_sub_matrix_m_calc}.
Also, for anisotropic material $n$ the eigenvector solutions in Eq.~\eqref{eq:2x2_eigenvector_V_U_biaxial}, given in Eqs.~\eqref{eq:2x2_eigenvector_V_biaxial_calc} and~\eqref{eq:2x2_eigenvector_U_biaxial_calc} with all the information inserted, deliver the matrices $\mathbf{E}_{ij,n}$ and $\mathbf{H}_{ij,n}$ for the anisotropic material $n$ in Eq.~\eqref{eq:refl_sub_matrix_n_calc}.
From there, it is explained in the Appendix how the reflection Müller-matrix $R_\textrm{Müller}$ in Eq.~\eqref{eq:jones_to_mueller} is finally calculated.\\

\section{Creating a database and fitting the theoretical model
\label{sec:database_fit}}
With the Müller-Stokes-formalism the expected reflected polarisation states represented by a Stokes-vector can be determined by multiplying the Müller-matrix of the reflecting sample $R_\textrm{Müller}$ with the Stokes-vector before the sample \cite{Tompkins2005}
\begin{equation}
    \vec{S} = \mathbf{R}_\textrm{Müller}(\sigma_1, \sigma_2, \varphi) \vec{S}_\textrm{init}.
    \label{eq:resulting_stokes}
\end{equation}
The initial Stokes-vector represents a $\ang{-45}$-polarised light beam and is multiplied by the Müller-matrix of a compensator with its slow axis oriented at $\ang{0}$, respectively. Thus, the initial polarisation state becomes \cite{Tompkins2005}
\begin{equation}
    \vec{S}_\mathrm{init} =
    \begin{pmatrix}
        1 & &0 &-\cos{\delta} &\sin{\delta}
    \end{pmatrix}^T,
    \label{eq:init_stokes}
\end{equation}
where $\delta$ is the retardation of the optional compensator resulting in $\cos (\delta=\pi/2) = 0$ and $\sin (\delta=\pi/2) = 1$ for a quarter wave plate compensator and $\cos (\delta=0) = 1$ and $\sin (\delta=0) = 0$ for no compensator. 
The latter is the case in this work.\\
The resulting Stokes vector $\vec{S}$ is calculated for each reflection Müller-matrix $\tensor{\mathbf{R}}(\sigma_1, \sigma_2, \varphi)$ of each stress state $(\sigma_1, \sigma_2, \varphi)$.
By that a database is created for the sweeps of $\sigma_1, \sigma_2$ between \SI{-15}{\mega\pascal} and \SI{15}{\mega\pascal} in \SI{1}{\mega\pascal} steps and between \SI{-5}{\mega\pascal} and \SI{5}{\mega\pascal} in \SI{0.2}{\mega\pascal} steps, while $\varphi$ is iterated between $\ang{0}$ and $\ang{90}$ in $\ang{1}$ steps.
Additionally, the database is generated for the geometry shown in Fig.~\ref{fig:index_ellipsoid} as well as for the plane of incidence turned by $\vartheta = \ang{-45}$ or the sample axes turned by \ang{45}, respectively.
The database is then separated into different signal types for the first two Stokes-vector components $\iota \in \{0, 1\}$ and the orientation of the plane of incidence $\vartheta \in \{\ang{0}, \ang{-45}\}$.
For each of the four signal types $S_{0/1,\ang{0}/\ang{-45}}(\sigma_1, \sigma_2, \varphi)$, those Stokes-vector components show a nearly linear relation to $\sigma_1$ and $\sigma_2$ for each fixed $\varphi$.
This behaviour can be described with
\begin{equation}
    S(\sigma_1, \sigma_2, \varphi) = A \sigma_1 + B \sigma_2 + c.
    \label{eq:flat_surface}
\end{equation}
For different azimuth angles $\varphi$ the slopes $A(\varphi)$ and $B(\varphi)$ follow opposing third-order polynomials of $\varphi$.
Hence, Eq.~\eqref{eq:flat_surface} is extended to a third order parametric family of $\varphi$
%\begin{widetext}
    \begin{equation}
    	\begin{aligned}
    		S(\sigma_1, \sigma_2, \varphi) =& (a + d \varphi^3 + e \varphi^2 + f \varphi) \sigma_1 \\
    		&+ (b - d \varphi^3 - e \varphi^2 - f \varphi)  \sigma_2 + c.
  		\end{aligned}
        \label{eq:flat_surface_trd_order_family}
    \end{equation}
%\end{widetext}
For each of the four signals this dependency on $\sigma_1$, $\sigma_2$ and $\varphi$ is fitted.
\begin{widetext}
    \begin{figure}[h]
        \centering
        \includegraphics[width=\textwidth]{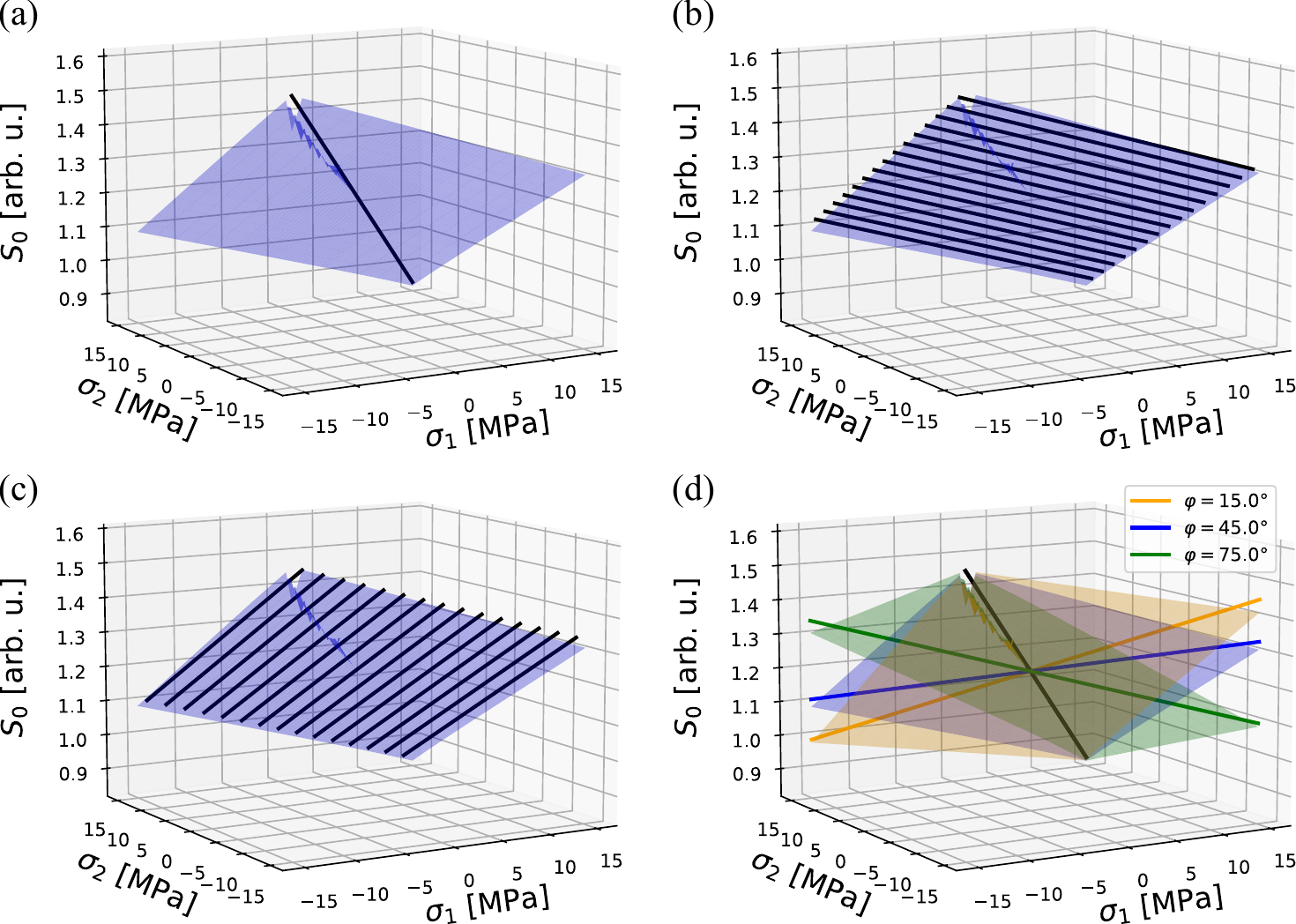}
        \captionsetup{width=\textwidth}
        \caption{For each azimuth angle $\varphi \in [\ang{5}, \ang{85}]$, the database includes a nearly flat surfaces representing the calculated Stokes vector component $S_0$ plotted against the stress component values $\sigma_1$ and $\sigma_2$.
        Here, in figures (a) to (c), the zeroth Stokes vector component is plotted $S_0$ against $\sigma_1$ and $\sigma_2$ for the azimuth angle $\varphi = \ang{45}$.
        Also, the results of the fits for the one dimensional profiles (a) along the diagonal to fit the signal axis intersect, (b) along the $x$-direction and (c) along the $y$-direction to fit the corresponding slopes are shown there.
        In panel~(d) $S_0$ is plotted against $\sigma_1$ and $\sigma_2$ for $\varphi \in \{\ang{15}, \ang{45}, \ang{75}\}$ together with the coinciding diagonal profiles and the diagonals perpendicular to the coinciding diagonal as indication of the fit results for each angle $\varphi$.
        All plots refer to the data of the zeroth Stokes vector component $S_0$ and the plane of incidence in the $xz$-plane with a $\ang{0}$ rotation of the sample orientation around the $z$-axis.}
        \label{fig:fit_procedure}
    \end{figure}
\end{widetext}
The fit procedure uses a series of one-dimensional fits along the dimensions $\sigma_1$, $\sigma_2$ and $\varphi$ instead of one three-dimensional fit in the following way:\\
First, one figures out the signal axis intersecting $c$ by linear fitting the profile line, where $\sigma_1 = \sigma_2$, for each azimuth angle $\varphi \in [\ang{5}, \ang{85}]$ considered in the database.
Since for $\sigma_1 = \sigma_2$ the azimuthal orientation is irrelevant, all surfaces for each azimuth intersect in this profile line, which allows a global fit of this profile line including the data points for each azimuth in order to achieve a more accurate results for the intersect.
The surface from the database and the fitted linear profile along the diagonal $\sigma_1 = \sigma_2$ are plotted in Fig.~\ref{fig:fit_procedure}(a).\\
Second, the slopes of the surface in $\sigma_1$- and $\sigma_2$-direction are determined by linear fitting the profiles for each fixed value of $\sigma_2$ and $\sigma_1$, respectively.
This is performed by using Eq.~\eqref{eq:flat_surface} for a fixed $\sigma_2$
\begin{equation}
    S(\sigma_1, \sigma_2=\mathrm{const.}, \varphi=\mathrm{const.}) = A \sigma_1 + B \sigma_2 + c = A \sigma_1 + c_B
    \label{eq:fitting_A}
\end{equation}
and for a fixed $\sigma_1$
\begin{equation}
    S(\sigma_1=\mathrm{const.}, \sigma_2, \varphi=\mathrm{const.}) = A \sigma_1 + B \sigma_2 + c = B \sigma_2 + c_A
    \label{eq:fitting_B}
\end{equation}
as fit functions of the profiles in order to obtain the slopes $A(\varphi)$ and $B(\varphi)$ for each $\varphi$.
Those profiles in $\sigma_1$-direction for several fixed $\sigma_2$ and in $\sigma_2$-direction for several fixed $\sigma_1$ are plotted on the surface plot for the azimuth angle $\varphi=\ang{45}$ in Fig.~\ref{fig:fit_procedure}(b) and~\ref{fig:fit_procedure}(c), respectively.
The slopes for each profile are averaged to one slope value along the $\sigma_1$-direction and one along the $\sigma_2$-direction.
That is done for all azimuths $\varphi \in [\ang{5}, \ang{85}]$, since for the angles near \ang{0} and \ang{90} the surface is no longer a differentiable linear function for the data calculated by the formalism.\\
For all fits in the described steps the data for $\sigma_1 = \sigma_2 < 0$ are ignored due to the singularities of the calculation results, visible in Fig.~\ref{fig:fit_procedure}.\\
Third, considering the resulting slopes for each $\varphi$ a dependency on $\varphi$ appears in the form of a third-order polynomial function.
Therefore, the $\varphi$ dependency of Eq.~\eqref{eq:flat_surface_trd_order_family} is determined using the results of $A(\varphi)$ and $B(\varphi)$ and the fit functions
\begin{equation}
    A(\varphi) = d \varphi^3 + e \varphi^2 + f \varphi + a
    \label{eq:A_of_phi}
\end{equation}
and
\begin{equation}
    B(\varphi) = -d \varphi^3 - e \varphi^2 - f \varphi + b,
    \label{eq:B_of_phi}
\end{equation}
respectively.
The $\varphi$-dependency of the surfaces in the database is depicted in Fig.~\ref{fig:fit_procedure}d for $\varphi \in \{\ang{15}, \ang{45}, \ang{75}\}$.
Furthermore, diagonal profiles are used as in Fig.~\ref{fig:fit_procedure}(a) as well as the profiles perpendicular to the coinciding diagonals are shown as lines to compare the fit results following Eq.~\eqref{eq:flat_surface_trd_order_family} to the database.
The perpendicular profiles to the diagonals refer to the results of the slopes $A(\varphi)$ and $B(\varphi)$ fitted with Eqs.~\eqref{eq:A_of_phi} and~\eqref{eq:B_of_phi}, respectively, indicating their $\varphi$-dependence.
That fit completes the parametrization of the family of surface functions $S_0(\sigma_1, \sigma_2, \varphi)$ in Eq.~\eqref{eq:flat_surface_trd_order_family}.
This procedure is performed for the first two Stokes vector components, in the detector's plane of incidence orientations of $\vartheta = \ang{0}$ and $\vartheta = \ang{-45}$.
Hence, these parametrized surface functions are used to determine the stress state $(\sigma_1, \sigma_2, \varphi)$ from the measured or calculated Stokes vector components.

\section{Quantification procedure
\label{sec:quantification}}
The fitting procedure using Eq.~\eqref{eq:flat_surface_trd_order_family} to describe the stress state $(\sigma_1, \sigma_2, \varphi)$ dependencies of any first or second Stokes-vector component signal achieves for the combination of all eight described signals a possibility to directly quantify the stress state.\\
This quantification is performed using a system of three equations of the form of Eq.~\eqref{eq:flat_surface_trd_order_family} for the three signals $S_{\kappa}(\sigma_1, \sigma_2, \varphi)$, $S_{\lambda}(\sigma_1, \sigma_2, \varphi)$ and $S_{\mu}(\sigma_1, \sigma_2, \varphi)$ which are chosen from four combinations of Stokes vector components $\left\{ S_{\iota,\vartheta}, \; \textrm{where} \; \vartheta \in \{\ang{0},\ang{-45}\}, \iota \in \{0, 1\}\right\}$ calculated for the database with the plane of incidence orientation $\vartheta$ to the $xz$-plane around the $z$-axis and the Stokes vector component $\iota$.
For each of these four combinations of three Stokes vector components, the system of the three equations, including the fitted parameters for each signal determined by the fitting procedure described above, is solved for $(\sigma_1, \sigma_2, \varphi)$ in the following manner:\\
First, Eq.~\eqref{eq:flat_surface_trd_order_family} for signal $\mu$ can be written as
%\begin{widetext}
    \begin{equation}
    	\begin{aligned}
        	0 = &(a_{\mu} + d_{\mu} \varphi^3 + e_{\mu} \varphi^2 + f_{\mu} \varphi) \sigma_1\\
        	&+ (b_{\mu} - d_{\mu} \varphi^3 - e_{\mu} \varphi^2 - f_{\mu} \varphi)  \sigma_2 + c_{\mu} - S_{\mu},
        \end{aligned}
        \label{eq:flat_surface_trd_order_family_mu}
    \end{equation}
%\end{widetext}
and its right hand side is calculated for each azimuth $\varphi$ considering the $\varphi$-dependent values for $\sigma_1$ and $\sigma_2$
These arise solving the Eqs.~\eqref{eq:flat_surface_trd_order_family} of the signals $\kappa$ and $\lambda$ for $\sigma_1$ and $\sigma_2$ considering each originating azimuth $\varphi$ from the database.
Since the absolute value of right-hand side of Eq.~\eqref{eq:flat_surface_trd_order_family_mu} is calculated for each originating azimuth $\varphi$ and the $\varphi$-dependent $\sigma_1$ and $\sigma_2$, the minimum of those is evaluated and the corresponding $\varphi$ is considered as the azimuth angle corresponding to the searched stress state causing the measured signal.\\
Second, the Eqs.~\eqref{eq:flat_surface_trd_order_family} of the signals $\kappa$ and $\lambda$ are solved for $\sigma_1$ and $\sigma_2$ once again, but this time only for the already determined $\varphi$.
Hence, one has determined the complete stress state $(\sigma_1, \sigma_2, \varphi)$, which has caused the Stokes vector components $S_{\kappa}$, $S_{\lambda}$ and $S_{\mu}$ representing the measurable signals.\\
The results of $\sigma_1$, $\sigma_2$ and $\varphi$ for each of the four combinations are averaged afterwards to improve the accuracy.

\section{Accuracy of the quantification by the fitted model
\label{sec:accuracy}}
\begin{figure}[h]
    \centering
    \includegraphics[width=0.45\textwidth]{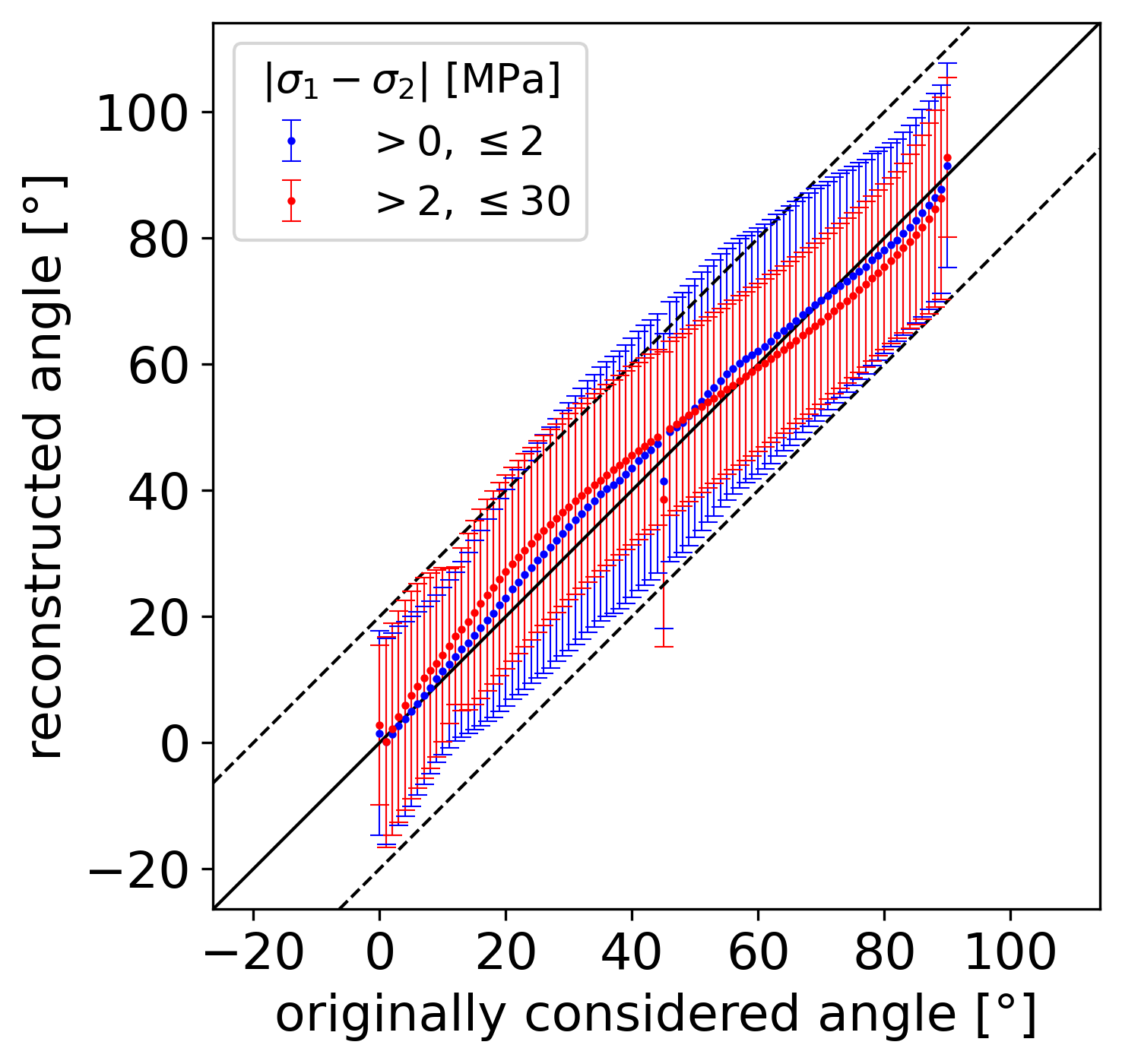}
    \caption{Determined azimuth angles $\varphi$ by the quantification of the single datasets plotted against the originating angles $\varphi$ of the corresponding dataset.
    The determined values are averaged and plotted with the standard deviation as error bars for datasets with the same originating angles.
    Also, the different colours represent the averaged angles separated in ranges of $|\sigma_1 - \sigma_2|$.
    Additionally, the identity line is plotted in solid black and the $\pm \ang{20}$ deviation lines in dashed black.
    \label{fig:angle_check}}
\end{figure}

In order to evaluate the accuracy of the fitting and quantification procedure, a self-verification is performed, where all the data sets in the database are quantified and the determined stress states are compared to the originating ones.
As explained in Section~\ref{sec:database_fit} on page~\pageref{sec:database_fit}, the database has been created and our model has been fitted to it, which is used for the quantification of each dataset, as described in Section~\ref{sec:quantification} on page~\pageref{sec:quantification}.
Using the resulting parametrized fits the stress state $(\sigma_1, \sigma_2, \varphi)$ is individually reconstructed from all four combinations of the stokes vector components $S_{\kappa}$, $S_{\lambda}$ and $S_{\mu}$, where $\kappa, \lambda, \mu$ corresponds to the index combinations $\iota,\vartheta$ with $\vartheta \in \{\ang{0},\ang{-45}\}$ and $\iota \in \{0, 1\}$, of each dataset.
The four resulting values of $\sigma_1$, $\sigma_2$ and $\varphi$ are averaged to obtain the final reconstructed stress state result for each dataset.
The thereby determined angles $\varphi$ are compared to the corresponding angles that have been originally considered to calculate each dataset of Stokes vectors.
That comparison is depicted as plot against each other in Fig.~\ref{fig:angle_check}.
The reconstructed angles are averaged for all datasets with the same originating angles, but separately for certain ranges of certain values of $|\sigma_1 - \sigma_2|$, e.g. lower stress component differences $\SI{0}{\mega\pascal} < |\sigma_1 - \sigma_2| \leq \SI{2}{\mega\pascal}$ and higher differences $\SI{2}{\mega\pascal} < |\sigma_1 - \sigma_2| \leq \SI{30}{\mega\pascal}$.
The error bars of the reconstructed angles are the standard deviations of their averages for each originating angle.
Additionally, the angles in the regions where the angle is expected to be $\varphi=\ang{0}$ or $\varphi=\ang{90}$, respectively, the reconstructed ones have been corrected to values modulo \ang{90} in order to reasonably average the values in those regions.
Admittedly, this step artificially creates better results for the averaged angles in those regions, since this procedure would not be possible for actually measured or calculated datasets without the reference to the originating stress state.\\
On the one hand, the determined angles by the quantification of datasets with stress differences $|\sigma_1 - \sigma_2| \leq 2$ show higher standard deviations than those with $|\sigma_1 - \sigma_2| > 2$, as visible in Fig.~\ref{fig:angle_check}.
That is actually caused by the small differences $|\sigma_1 - \sigma_2| \leq 2$ since the angles for datasets with $|\sigma_1 - \sigma_2| = 0$ are not definitely determinable, as the stress or index ellipses are circles.
Consequently, the data sets with stress component differences close to zero are more error-prone.\\
On the other hand, the standard deviations of the other range $|\sigma_1 - \sigma_2| > 2$ are around \ang{15}, while all means deviate less than half the standard deviation from the identity line, indicating a decisively better accuracy only degraded by outliers.
The outliers originate from those datasets where the fitted linear dependencies differ most from the slightly parabolic surfaces in Fig.~\ref{fig:fit_procedure}, that is, those data points far from the intersecting diagonal.\\
For both difference ranges in Fig.~\ref{fig:angle_check} a conspicuous deviation of the averaged reconstructed angle from the identity line and the overall trend reconstructed angles appears at $\varphi = \ang{45}$ as well as similar but less distinct at $\varphi = \ang{0}$ and $\varphi = \ang{90}$.
That is presumably caused by the fact that at $\varphi = \ang{0}$ and $\varphi = \ang{90}$ the surfaces in the $S_i$-$\sigma_1$-$\sigma_2$-spaces are not linear for the plane of incidence orientation of $\vartheta = \ang{0}$ but show a sharp bend. The same is the case for the orientation of the principal axis $\varphi = \ang{45}$ and the orientation plane of incidence of $\vartheta = \ang{-45}$\\

\begin{figure}[h]
    \centering
    \includegraphics[width=0.45\textwidth]{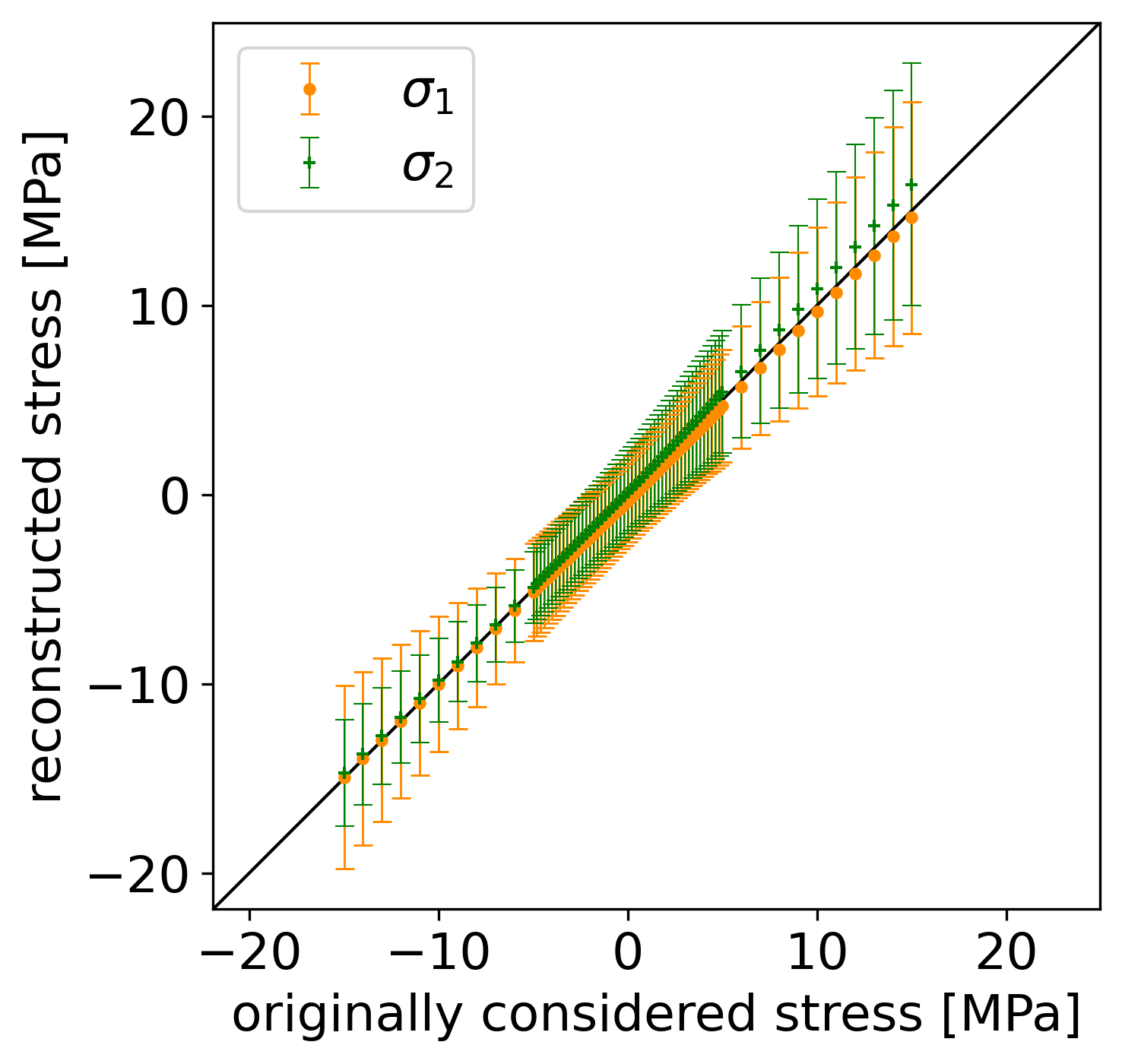}
    \caption{Determined stress values $\sigma_1$ and $\sigma_2$ by the quantification of the single datasets plotted against the originating stress value of the corresponding dataset.
    The determined values are averaged and plotted with the standard deviation as error bars for datasets with the same originating angles.
    The point density of the data in the region between \SI{-5}{\mega\pascal} and \SI{5}{\mega\pascal} is too high to visualise the data points separated from each other.
    Additionally, the identity line is plotted in black.
    \label{fig:linearity-check}}
\end{figure}

The determined values of $\sigma_1$ and $\sigma_2$ are calculated by the previously determined angles $\varphi$ of the corresponding datasets.
In Fig.~\ref{fig:linearity-check} the averages of those determined stress values with their standard deviations as error bars are plotted against the originally considered in the corresponding datasets.
In Fig.~\ref{fig:linearity-check} appears the decreasing standard deviations for smaller absolute stress values.
Furthermore, the determined averages deviate more from the identity line for positive than for negative stresses.
Both effects are caused by the accuracy of the surface fit functions, which differ less from the calculated data surfaces in Fig.~\ref{fig:fit_procedure} for small stress values and, also, typically follow rather the parabolic arm of the surface's negative stresses region than that of the positive one caused by the procedure of fitting.\\
The overall accuracy of the quantification procedure, including the fitting procedure, lies below $\Delta \sigma \leq \SI{5}{\mega\pascal}$ for stresses $\SI{-5}{\mega\pascal} \leq \sigma_{1/2} \leq \SI{5}{\mega\pascal}$ and below $\Delta \sigma \leq \SI{8}{\mega\pascal}$ for stresses $\SI{-15}{\mega\pascal} \leq \sigma_{1/2} \leq \SI{15}{\mega\pascal}$, while the angle $\varphi$ is determined with an accuracy $\Delta \varphi \leq \ang{20}$.
For differences of stress values above $|\sigma_1 - \sigma_2| > \SI{2}{\mega\pascal}$, the corresponding angle $\varphi$ is determined more accurately than $\Delta \varphi \leq \ang{15}$.

\section{Experimental approaches
\label{sec:exp_approaches}}
\begin{figure}[h]
	\centering
	\includegraphics[width=0.45\textwidth]{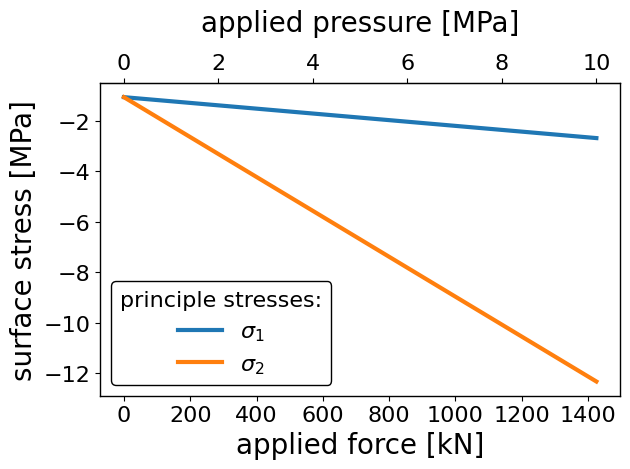}
	\caption{Principle axes of stress calculated from linear fits of SCALP measurements in $x$- and $y$-direction depending on an applied compressive force along the $y$-direction.
	\label{fig:scalp_measurement}}
\end{figure}

Surface stress measurements using the SCALP are often simultaneously performed to four-point bending tests, for example by Aben et. al. \cite{Aben2010} or more recently by Efferz et. al. \cite{Efferz2024}.
In contrast, we developed a forcing apparatus applying stress along one of the surface plane directions in order to recreate parts of the simulated data experimentally, e.g. these profiles of the surface plots in Fig.~\ref{fig:fit_procedure} where either $\sigma_1 = 0$ or $\sigma_2 = 0$.
Measurements with the reflection polarimeter, whose results are theoretical calculated by Müller-Stokes formalism here, are in a developmental stage.
However, SCALP measurements have been performed as a reference to validate the forcing apparatus and the linearity of those measurements our model depends on.
For a compressive forcing along the $y$-direction we have taken SCALP measurements along the $x$- and $y$-direction (corresponding to \ang{0} and \ang{90}), linearly fit them and calculated the principle axes of stress $\sigma_1 = \sigma_x$ and $\sigma_2 = \sigma_y$, according to \cite{Chen2013}, depending on the applied compressive force.
The linear dependencies are shown in Fig.~\ref{fig:scalp_measurement}.
The first principle axis $\sigma_1$ corresponding to the $x$-direction is nearly independent of the forcing compared to the stress along the second principle axis $\sigma_2$ corresponding to the $y$-direction, which is the direction of the applied force.
The measured surface stress $\sigma_1$ corresponds with an offset of $\approx \SI{1}{\mega\pascal}$ pretty close to the applied pressure.
This indicates the validity of our forcing apparatus.
Hence, we will be able to experimentally recreate the used theoretical database by our reflection polarimeter measurements in the next step.

\section{Conclusion
\label{sec:conclusion}}
In conclusion, we showed that a direct quantification of the stress state $(\sigma_1, \sigma_2, \varphi)$ is possible for our simplified fit model of the theoretical data, only limited by the accuracy defined as the standard deviation of $\Delta \sigma \leq \SI{5}{\mega\pascal}$ for stresses $\SI{-5}{\mega\pascal} \leq \sigma_{1/2} \leq \SI{5}{\mega\pascal}$.
In comparison, this determination error is higher than the error of the SCALP, which is the widely used state-of-the-art instrument for surfaces stress measurements.
In the manufacturer's data on the GlasStress website \cite{GlasStress2024}, the SCALP measurement errors are specified with $\Delta \sigma < \SI{1}{\mega\pascal}$ for $\sigma < \SI{4}{\mega\pascal}$ and 5\% for $\sigma \geq \SI{4}{\mega\pascal}$.
However, our method allows for a complete determination of the stress state under the geometric conditions described in Section~\ref{sec:special_stress_geometry} on page~\pageref{sec:special_stress_geometry} and Fig.~\ref{fig:index_ellipsoid} with an error for the azimuth angle of at maximum $\Delta \varphi \leq \ang{20}$.
Also, for higher absolute stress values, the error, with which the stress values $\SI{-15}{\mega\pascal} \leq \sigma_{1/2} \leq \SI{15}{\mega\pascal}$ are determined, increases to $\Delta \sigma \leq \SI{8}{\mega\pascal}$.
That is not too far away from the state of the art, where one has to consider, e.g. in case of the SCALP, that one only probes the stress in one direction and not the complete stress state.\\
Our work is considering the self-verification of the quantitative stress state reconstruction of theoretically calculated Stokes vector components.
That means that there are two differences to the practical case.
The practical case would use a measured database instead of a calculated one to quantitatively reconstruct an unknown sample of the same material used for the database recording.
The first difference would be the precision of the measurement and the experimental errors, which will have an influence on the overall accuracy of the database fits and thus the reconstruction.
The second difference is the fact that an unknown sample is probed. 
Reconstructing the azimuth for azimuths near \ang{0} and \ang{90} can end up in a situation where the principal stresses $\sigma_1$ and $\sigma_2$ could be confused corresponding to a false assignment of $\varphi$ to \ang{0} instead of \ang{90} or the other way around.
That would not be recognized by the corresponding actual azimuth and therefore would not be prevented by using the modulo \ang{90} correction as described for the self-verification in Section~\ref{sec:accuracy} on page~\pageref{sec:accuracy}.
Ultimately, this could cause an interchange of the $\sigma_1$ and $\sigma_2$ values that could not be identified.\\
Nevertheless, our method paves the way for quantitative reconstruction of the surface stress state by a polarimetric measurement on dielectric materials using a reference database of the same material as probed.
We showed for our model a theoretical precision of at least the same order of magnitude as the state-of-the-art instrument for surface stress measurements with the extension of being able to quantify the complete stress state including both stress values of the principal axes in the surface plane as well as their azimuthal orientation in the plane.\\
Of course, the model has its limitations to dielectric material and positions of it, where the proposed geometry is present.
Furthermore, the model is designed for an idealised case of no impurities in the dielectric medium or the atmosphere causing scattering effects.
In most cases of float glass or tempered glass applications the occurrence of impurities and their disturbing effects should neglectable small.\\
While the considered surface stress range of \SIrange{-15}{15}{\mega\pascal} rather covers the laminated glasses used as automotive windshields, thermally tempered glass with residual surface stresses in the range of \SIrange{-80}{-120}{\mega\pascal} \cite{Chen2013} is typically used for automotive side windows and is closer to the extreme case of glass rupture.
One can imagine by extrapolating the trend of the rising errorbars in Fig.~\ref{fig:linearity-check} a decreasing precision for measurements of that high absolute surface stress values.
This is again intrinsically caused by the actually parabolic behaviour of the surfaces in Fig.~\ref{fig:fit_procedure} and cannot be solved by a linear description of the total stress range.
However, one could think about a local linear description in the regions of interest.
For thermally tempered glass this would be, for example, a range of \SIrange{-80}{-120}{\mega\pascal}.
Since this is a range of \SI{40}{\mega\pascal}, one would expect similar accuracies as for the range of \SI{30}{\mega\pascal} evaluated here.

\section{Outlook
\label{sec:outlook}}
For the theoretical calculations and the stress determination procedure, efforts are being made to extend the formalism to a generally absorbing material by including the imaginary part in the complex refractive index representing the absorption in the equations.
On the experimental side, measurements are currently being prepared to test the same fitting procedure for a measured database.
In that case, the database is measured by externally applying a mechanical stress on glass panes in a single direction for several angles $\tilde{\varphi} \in [\ang{0}, \ang{180}]$ thus imitating a stressed glass state $(\sigma_1, \sigma_2=0, \varphi)$ and $(\sigma_1=0, \sigma_2, \varphi)$ for $\varphi \in [\ang{0}, \ang{90}]$.
These data can be extrapolated to complete surfaces $S(\sigma_1, \sigma_2, \varphi=const.)$ in the linear regions of the relations $S(\sigma_1, \sigma_2=0, \varphi=const.)$ and $S(\sigma_1=0, \sigma_2, \varphi=const.)$. With the same procedure, as described in this work, self-verification can be performed to compare the theoretically possible and experimentally achievable precision of the quantification method.\\
Moreover, the determination of certain parameters, for example, the stress state $(\sigma_1, \sigma_2, \varphi)$, using a database containing datasets with other parameters, such as Stokes vector components, is a typical application for machine learning. We plan to feed machine learning algorithms with the databases of measured or calculated signal datasets in order to evaluate the stress states replacing the manual numerical fitting and solving procedure.
Probably, the accuracy of the quantitative determination of the stress states can be improved by avoiding the restriction of fitting a linear model and using a "semi-analytical" solution of the resulting system of equations.

\begin{acknowledgments}
We acknowledge the Federal Ministry of Education and Research (Bundesministerium für Bildung und Forschung: BMBF) of the Federal Republic of Germany for funding the project (Grant No. 13N15888).
We express further thanks to our project partners Volkswagen A.G., Saint-Gobain Sekurit Deutschlang GmbH and Schmidt + Haensch GmbH \& Co. as well as to Prof. A. Egner for fruitful discussions. Special thanks go to Prof. I. Abdulhalim for personal communications concerning his paper on the $2\times2$-matrix formalism \cite{Abdulhalim1999}.
\end{acknowledgments}

\begin{widetext}

\appendix*
\section*{Appendix
\label{sec:appendix}}

We provide the following calculations as supplemental material in order to retrace the intermediate steps between the given equations in the paper.\\
For air or vacuum as the medium $m$ the matrices in Eq.~\eqref{eq:gen_refl_sub_matrix} become
\begin{subequations}
    \begin{eqnarray}
        \mathbf{E}_{12m} &=& \mathbf{E}_{34m} =
        \begin{pmatrix}
            1   &0  \\
            0   &1
        \end{pmatrix},\\
        \mathbf{H}_{12m} &=&
        \begin{pmatrix}
            \mathbf{Q}_{1m} 
            \begin{pmatrix}
                1   \\
                0
            \end{pmatrix}& \mathbf{Q}_{2m} 
            \begin{pmatrix}
                0   \\
                1
            \end{pmatrix} 
        \end{pmatrix} \nonumber \\ &=&
        \begin{pmatrix}
            0 &-\nu_{z2m}\\
            \nu_{z1m} - b_{x1m} \nu_{x1m} &0
        \end{pmatrix} \nonumber \\ &=&
        \begin{pmatrix}
            0 &- \cos{\gamma_m}\\
            \left( \cos{\gamma_m} - \frac{sin^2{\gamma_m}}{\cos{\gamma_m}} \right) &0
        \end{pmatrix} 
        \quad \textrm{and} \\
        \mathbf{H}_{34m} &=&
        \begin{pmatrix}
            \mathbf{Q}_{3m} 
            \begin{pmatrix}
                1   \\
                0
            \end{pmatrix}& \mathbf{Q}_{4m} 
            \begin{pmatrix}
                0   \\
                1
            \end{pmatrix} 
        \end{pmatrix} \nonumber \\ &=&
        \begin{pmatrix}
            0 &-\nu_{z4m}\\
            \nu_{z3m} - b_{x3m} \nu_{x3m} &0
        \end{pmatrix} \nonumber \\ &=&
        \begin{pmatrix}
            0 &\cos{\gamma_m}\\
            - \left( \cos{\gamma_m} - \frac{sin^2{\gamma_m}}{\cos{\gamma_m}} \right) &0
        \end{pmatrix} 
    \end{eqnarray}
    \label{eq:refl_sub_matrix_m_calc}
\end{subequations}
by inserting the results of Eqs.~\eqref{eq:2x2_eigenvector_V_air} and \eqref{eq:2x2_eigenvector_U_air} together with Eqs.~\eqref{eq:U-V-dependence} and~\eqref{eq:rel_psi_e_h_simplified}.\\
From Eqs.~\eqref{eq:2x2_eigenvector_V_biaxial} and~\eqref{eq:2x2_eigenvector_U_biaxial} with all inserted information calculated in Section~\ref{sec:special_stress_geometry} follows
% \begin{widetext}%%%
    \begin{subequations}
        \begin{eqnarray}
            V_{e1n} &=& N^{-1}
            \begin{pmatrix}
                1\\-\frac{g_{11}}{g_{12}}
            \end{pmatrix}
            = N^{-1}
            \begin{pmatrix}
                1\\\frac{\varepsilon_{xy}}{\nu_{xn}^2 + \nu_{z1n}^2 - \varepsilon_{yy}}
            \end{pmatrix}
            \nonumber \\ &=& N^{-1}
            \begin{pmatrix}
                1\\\frac{\sin(\varphi - \vartheta) \cos(\varphi - \vartheta) (n_1^2 - n_2^2)}{\nu_{xn}^2 + \nu_{z1n}^2 - \cos^2(\varphi - \vartheta) n_2^2 - \sin^2(\varphi - \vartheta) n_1^2}
            \end{pmatrix} \\
            \textrm{with} \quad
            N &=& \sqrt{1 + \left( \frac{g_{11}}{g_{12}} \right)^2}
            = \sqrt{1 + \left( \frac{-\varepsilon_{xy}}{\nu_{xn}^2 + \nu_{z1n}^2 - \varepsilon_{yy}} \right)^2} \nonumber \\
            &=& \sqrt{1 + \left( \frac{\sin(\varphi - \vartheta) \cos(\varphi - \vartheta) (n_1^2 - n_2^2)}{\nu_{xn}^2 + \nu_{z1n}^2 - \cos^2(\varphi - \vartheta) n_2^2 + \sin^2(\varphi - \vartheta) n_1^2} \right)^2}
        \end{eqnarray}
        \label{eq:2x2_eigenvector_V_biaxial_calc}
    \end{subequations}
% \end{widetext}%%%
and
% \begin{widetext}%%%
    \begin{subequations}
        \begin{eqnarray}
            U_{e2n} &=& M^{-1}
            \begin{pmatrix}
                -\frac{g_{22}}{g_{21}}\\1
            \end{pmatrix}
            = M^{-1}
            \begin{pmatrix}
                \frac{-\varepsilon_{xy}}{\varepsilon_{xx} + b_{x2n} \nu_{xn} \nu_{z2n} - \nu_{z2n}^2}\\1
            \end{pmatrix}
            \nonumber \\ &=& M^{-1}
            \begin{pmatrix}
                \frac{\sin(\varphi - \vartheta) \cos(\varphi - \vartheta) (n_2^2 - n_1^2)}{\cos^2(\varphi - \vartheta) n_2^2 + \sin^2(\varphi - \vartheta) n_2^2 + b_x \nu_{xn} \nu_{zn} - \nu_{zn}^2}\\1
            \end{pmatrix} \\
            \textrm{with} \quad
            M &=& \sqrt{1 + \left( \frac{g_{22}}{g_{21}} \right)^2}
            = \sqrt{1 + \left( \frac{\varepsilon_{xy}}{\varepsilon_{xx} + b_x2n \nu_{xn} \nu_{z2n} - \nu_{z2n}^2} \right)^2} \nonumber\\
            &=& \sqrt{1 + \left( \frac{\sin(\varphi - \vartheta) \cos(\varphi - \vartheta) (n_2^2 - n_1^2)}{\cos^2(\varphi - \vartheta) n_2^2 + \sin^2(\varphi - \vartheta) n_2^2 + b_x2n \nu_{xn} \nu_{z2n} - \nu_{z2n}^2} \right)^2}. \quad \quad 
        \end{eqnarray}
        \label{eq:2x2_eigenvector_U_biaxial_calc}
    \end{subequations}
% \end{widetext}%%%
Thus, Eq.~\eqref{eq:gen_refl_sub_matrix} for the anisotropic medium $n$ becomes
% \begin{widetext}%%%
    \begin{subequations}
        \begin{eqnarray}
            \mathbf{E}_{12n} &=&
            \begin{pmatrix}
                V_{e1,n} &U_{e2,n}
            \end{pmatrix} =
            \begin{pmatrix}
                N^{-1}  & -M^{-1}\frac{g_{2n,22}}{g_{2n,21}}   \\
                -N^{-1}\frac{g_{1n,11}}{g_{1n,12}}  &M^{-1}
            \end{pmatrix} \quad \textrm{and} \\
            \mathbf{H}_{12n} &=&
            \begin{pmatrix}
                V_{h1n} &U_{h2n}
            \end{pmatrix} =
            \begin{pmatrix}
                \mathbf{Q}_{1n}
                \begin{pmatrix}
                    N^{-1}\\ -N^{-1} \frac{g_{1n,11}}{g_{1n,12}}
                \end{pmatrix}& 
                \mathbf{Q}_{2n}
                \begin{pmatrix}
                    -M^{-1}\frac{g_{2n,22}}{g_{2n,21}}   \\M^{-1}
                \end{pmatrix}
            \end{pmatrix} \nonumber\\&=&
            \begin{pmatrix}
                \nu_{z1n} N^{-1} \frac{g_{1n,11}}{g_{1n,12}}   &-\nu_{z2n} M^{-1} \\
                \left( \nu_{z1n} - b_{x1n} \nu_{xn} \right) N^{-1}   &-\left( \nu_{z2n} - b_{x2n} \nu_{xn} \right) M^{-1}\frac{g_{2n,22}}{g_{2n,21}}
            \end{pmatrix}.
        \end{eqnarray}
        \label{eq:refl_sub_matrix_n_calc}
    \end{subequations}
% \end{widetext}%%%
With the results in Eqs.~\eqref{eq:refl_sub_matrix_m_calc} and~\eqref{eq:refl_sub_matrix_n_calc} the reflection matrix \eqref{eq:gen_refl-matrix_2x2} becomes
% \begin{widetext}%%%
    \begin{equation}
        \mathbf{R}_{mn} 
        = \left( \mathbf{E}_{12m}^{-1} \mathbf{E}_{34m} - \mathbf{H}_{12m}^{-1} \mathbf{H}_{34m} \right) \left( \mathbf{H}_{12n}^{-1} \mathbf{H}_{12m} - \mathbf{E}_{12n}^{-1} \mathbf{E}_{12m} \right)
        = \left( \mathbb{I}_2 - \mathbf{H}_{12m}^{-1} \mathbf{H}_{34m} \right) \left( \mathbf{H}_{12n}^{-1} \mathbf{H}_{12m} - \mathbf{E}_{12n}^{-1} \right)
        \label{eq:refl-matrix_2x2}
    \end{equation}
% \end{widetext}%%%
for the anisotropic medium $n$ and the medium $m$ being air or vacuum.
The reflection matrix $R_{mn}$ is related to the mode basis, i.e. the matrix elements represent the transitions between the corresponding modes \cite{Abdulhalim1999}.
For the used geometry with the plane of incidence in the $xz$-plane the transformation from the $xy$ field component basis to the ps-basis in \cite{Abdulhalim1999} simplifies to
% \begin{widetext}%%%
    \begin{equation}
        \begin{pmatrix}
            E_x\\ E_y
        \end{pmatrix}_{j,m/n} =
        \mathbf{T}_{j,m/n}
        \begin{pmatrix}
            E_p\\ E_s
        \end{pmatrix}_{m/n}=
        \begin{pmatrix}
            \cos{\gamma_{j,m/n}} &0 \\0 &1
        \end{pmatrix}
        \begin{pmatrix}
            E_p\\ E_s
        \end{pmatrix}_{m/n}
    \end{equation}
% \end{widetext}%%%
for mode type $j \in \{1,2\}$, which corresponds to $j \in \{3, 4\}$ in case of the reflected beam.
Thus, the the reflection matrix transforms to the sp-basis with
\begin{eqnarray}
    \tilde{R} &=& \mathbf{T}_{jm}^{-1} \mathbf{R}_{mn} \mathbf{T}_{jm} =  \mathbf{T}_{m}^{-1} \mathbf{R}_{mn} \mathbf{T}_{m}\nonumber \\ &=& 
    \begin{pmatrix}
        \frac{1}{\cos{\gamma_{m}}} &0 \\0 &1
    \end{pmatrix}
    \begin{pmatrix}
        r_{11} &r_{12} \\r_{21} &r_{22}
    \end{pmatrix}
    \begin{pmatrix}
        \cos{\gamma_{m}} &0 \\0 &1
    \end{pmatrix} \nonumber \\&=&
    \begin{pmatrix}
        r_{11} &\frac{r_{12}}{\cos{\gamma_{m}}} \\r_{21} \cos{\gamma_{m}} &r_{22}
    \end{pmatrix},
    \label{eq:refl_matrix_trafo_xy_to_sp}
\end{eqnarray}
since in the isotropic medium all modes' $j$ angles are $\gamma_{jm} = \gamma_m$, as shown in Fig.~\ref{fig:reflection_k}.
To transfer the reflection matrix from the Jones- to the Müller-Stokes-formalism \cite{MuellerMatrixInProceedings1948} the relation 
% \begin{widetext}%%%
    \begin{eqnarray}
        \mathbf{R}_\textrm{Müller} &=& A \left( \tilde{R} \otimes \tilde{R}^* \right) A^{-1} \nonumber \\&=&
        \begin{pmatrix}
            1   &0  &0  &1  \\
            1   &0  &0  &-1  \\
            0   &1  &1  &0  \\
            0   &i  &-i &0
        \end{pmatrix}
        \begin{pmatrix}
            r_{pp} \tilde{R}^*  &r_{ps} \tilde{R}^*  \\
            r_{sp} \tilde{R}^*  &r_{ss} \tilde{R}^*
        \end{pmatrix}
        \begin{pmatrix}
            \frac{1}{2}   &\frac{1}{2}  &0  &0  \\
            0   &0  &\frac{1}{2}  &-\frac{i}{2}  \\
            0   &0  &\frac{1}{2}  &\frac{i}{2}  \\
            \frac{1}{2}   &-\frac{1}{2}  &0 &0
        \end{pmatrix}
        \nonumber \\&=& \frac{1}{2}
        \begin{pmatrix}
            1   &0  &0  &1  \\
            1   &0  &0  &-1  \\
            0   &1  &1  &0  \\
            0   &i  &-i &0
        \end{pmatrix}
        \begin{pmatrix}
            r_{pp} \bar{r}_{pp} &r_{pp} \bar{r}_{sp}    &r_{ps} \bar{r}_{pp}    &r_{ps} \bar{r}_{sp}    \\
            r_{pp} \bar{r}_{ps} &r_{pp} \bar{r}_{ss}    &r_{ps} \bar{r}_{ps}    &r_{ps} \bar{r}_{ss}    \\
            r_{sp} \bar{r}_{pp} &r_{sp} \bar{r}_{sp}    &r_{ss} \bar{r}_{pp}    &r_{ss} \bar{r}_{sp}    \\
            r_{sp} \bar{r}_{ps} &r_{sp} \bar{r}_{ss}    &r_{ss} \bar{r}_{ps}    &r_{ss} \bar{r}_{ss}
        \end{pmatrix}
        \begin{pmatrix}
            1   &1  &0  &0  \\
            0   &0  &1  &-i  \\
            0   &0  &1  &i  \\
            1   &-1  &0 &0
        \end{pmatrix} \nonumber \\
        &=& \frac{1}{2}
        \begin{pmatrix}
            1   &0  &0  &1  \\
            1   &0  &0  &-1  \\
            0   &1  &1  &0  \\
            0   &i  &-i &0
        \end{pmatrix}
        \begin{pmatrix}
            r_{pp} \bar{r}_{pp} + r_{ps} \bar{r}_{sp}
            &r_{pp} \bar{r}_{pp} - r_{ps} \bar{r}_{sp}
            &r_{pp} \bar{r}_{sp} + r_{ps} \bar{r}_{pp}
            &i \left(-r_{pp} \bar{r}_{sp} + r_{ps} \bar{r}_{pp} \right)\\
            r_{pp} \bar{r}_{ps} + r_{ps} \bar{r}_{ss}
            &r_{pp} \bar{r}_{ps} - r_{ps} \bar{r}_{ss}
            &r_{pp} \bar{r}_{ss} + r_{ps} \bar{r}_{ps}
            &i \left(-r_{pp} \bar{r}_{ss} + r_{ps} \bar{r}_{ps} \right)\\
            r_{sp} \bar{r}_{pp} + r_{ss} \bar{r}_{sp}
            &r_{sp} \bar{r}_{pp} - r_{ss} \bar{r}_{sp}
            &r_{sp} \bar{r}_{sp} + r_{ss} \bar{r}_{pp}
            &i \left(-r_{sp} \bar{r}_{sp} + r_{ss} \bar{r}_{pp} \right)\\
            r_{sp} \bar{r}_{ps} + r_{ss} \bar{r}_{ss}
            &r_{sp} \bar{r}_{ps} - r_{ss} \bar{r}_{ss}
            &r_{sp} \bar{r}_{ss} + r_{ss} \bar{r}_{ps}
            &i \left(-r_{sp} \bar{r}_{ss} + r_{ss} \bar{r}_{ps} \right)
        \end{pmatrix}
        \nonumber \\&=&\frac{1}{2} \left(
        \begin{matrix}
            r_{pp} \bar{r}_{pp} + r_{ps} \bar{r}_{sp} +  r_{sp} \bar{r}_{ps} + r_{ss} \bar{r}_{ss}
            &r_{pp} \bar{r}_{pp} - r_{ps} \bar{r}_{sp} + r_{sp} \bar{r}_{ps} - r_{ss} \bar{r}_{ss}\\
            r_{pp} \bar{r}_{pp} + r_{ps} \bar{r}_{sp} -  r_{sp} \bar{r}_{ps} - r_{ss} \bar{r}_{ss}
            &r_{pp} \bar{r}_{pp} - r_{ps} \bar{r}_{sp}- r_{sp} \bar{r}_{ps} + r_{ss} \bar{r}_{ss}\\
            r_{pp} \bar{r}_{ps} + r_{ps} \bar{r}_{ss} + r_{sp} \bar{r}_{pp} + r_{ss} \bar{r}_{sp}
            &r_{pp} \bar{r}_{ps} - r_{ps} \bar{r}_{ss} + r_{sp} \bar{r}_{pp} - r_{ss} \bar{r}_{sp}\\
            i \left( r_{pp} \bar{r}_{ps} + r_{ps} \bar{r}_{ss} - r_{sp} \bar{r}_{pp} - r_{ss} \bar{r}_{sp} \right)
            &i \left( r_{pp} \bar{r}_{ps} - r_{ps} \bar{r}_{ss} - r_{sp} \bar{r}_{pp} + r_{ss} \bar{r}_{sp} \right)
        \end{matrix} \right. \nonumber \\&& \quad \quad \left.
        \begin{matrix}
            r_{pp} \bar{r}_{sp} + r_{ps} \bar{r}_{pp} + r_{sp} \bar{r}_{ss} + r_{ss} \bar{r}_{ps}
            &i \left(-r_{pp} \bar{r}_{sp} + r_{ps} \bar{r}_{pp} - r_{sp} \bar{r}_{ss} + r_{ss} \bar{r}_{ps} \right)\\
            r_{pp} \bar{r}_{sp} + r_{ps} \bar{r}_{pp} - r_{sp} \bar{r}_{ss} - r_{ss} \bar{r}_{ps}
            &i \left(-r_{pp} \bar{r}_{sp} + r_{ps} \bar{r}_{pp} + r_{sp} \bar{r}_{ss} - r_{ss} \bar{r}_{ps} \right)\\
            r_{pp} \bar{r}_{ss} + r_{ps} \bar{r}_{ps} + r_{sp} \bar{r}_{sp} + r_{ss} \bar{r}_{pp}
            &i \left(-r_{pp} \bar{r}_{ss} + r_{ps} \bar{r}_{ps} - r_{sp} \bar{r}_{sp} + r_{ss} \bar{r}_{pp} \right)\\
            i \left( r_{pp} \bar{r}_{ss} + r_{ps} \bar{r}_{ps} - r_{sp} \bar{r}_{sp} - r_{ss} \bar{r}_{pp} \right)
            &r_{pp} \bar{r}_{ss} - r_{ps} \bar{r}_{ps} - r_{sp} \bar{r}_{sp} + r_{ss} \bar{r}_{pp}
        \end{matrix} \right)
        \label{eq:jones_to_mueller}
    \end{eqnarray}
% \end{widetext}%%%
is used, where $\tilde{R}^*$ is the conjugate transpose of $\tilde{R}$ \cite{MuellerMatrixInProceedings1948}.

\end{widetext}

\bibliography{references}

\end{document}